\documentstyle[12pt,aaspp4]{article}
\begin{document}

\title{Complete and Simultaneous Spectral Observations of the
Black-Hole X-ray Nova XTE~J1118+480}

\author{J. E. McClintock\altaffilmark{1},
C. A. Haswell\altaffilmark{2},
M. R. Garcia\altaffilmark{1}, J. J. Drake\altaffilmark{1},
R. I. Hynes\altaffilmark{3}, H.~L.~Marshall\altaffilmark{4},
M.~P.~Muno\altaffilmark{4}, S. Chaty\altaffilmark{2},
P. M. Garnavich\altaffilmark{5}, P. J. Groot\altaffilmark{1},
W.~H.~G.~Lewin\altaffilmark{4},
C.~W.~Mauche\altaffilmark{6}, J. M. Miller\altaffilmark{4},
G. G. Pooley\altaffilmark{7}, C. R. Shrader\altaffilmark{8},
S. D. Vrtilek\altaffilmark{1}}

\small

\begin{abstract}
The X-ray nova XTE J1118+480 suffers minimal extinction (b = 62$^{\rm
o}$) and therefore represents an outstanding opportunity for
multiwavelength studies.  Hynes et al. (2000) conducted the first such
study, which was centered on 2000 April~8 using UKIRT, {\it EUVE}, {\it HST} and
{\it RXTE}.  On 2000 April 18, the {\it Chandra X-ray Observatory} obtained
data coincident with a second set of observations using all of these
same observatories.  A 30~ks grating observation using {\it Chandra} yielded a
spectrum with high resolution and sensitivity covering the range
0.24--7 keV. Our near-simultaneous observations cover $\approx$ 80\%
of the electromagnetic spectrum from the infrared to hard X-rays. The
UV/X-ray spectrum of XTE J1118+480 consists of two principal
components.  The first of these is a $\approx$ 24 eV thermal component
which is due to an accretion disk with a large inner disk radius:
$\gtrsim~35$R$_{\rm~Schw}$.  The second is a quasi power-law component
that was recorded with complete spectral coverage from 0.4--160~keV.
A model for this two-component spectrum is presented in a companion
paper by Esin et al. (2001).
\end{abstract}

\vskip -.15in


\footnotesize

\altaffilmark{1}{Harvard-Smithsonian Center for Astrophysics, 60
Garden Street, Cambridge, MA 02138; jem@cfa.harvard.edu,
mgarcia@cfa.harvard.edu, jdrake@cfa.harvard.edu,
pgroot@cfa.harvard.edu, svrtilek@cfa.harvard.edu}

\vskip -.1in

\altaffilmark{2}{Department of Physics and Astronomy, The Open
University, Walton Hall, Milton Keynes, MK7 6AA, UK;
c.a.haswell@open.ac.uk, s.chaty@open.ac.uk}

\vskip -.1in

\altaffilmark{3}{Department of Physics and
Astronomy, University of Southampton, Southampton, SO17, 1BJ;
rih@astro.soton.ac.uk} 

\vskip -.1in

\altaffilmark{4}{Center for Space Research, MIT, Cambridge, MA 02139;
hermanm@space.mit.edu, muno@space.mit.edu, lewin@space.mit.edu,
jmm@space.mit.edu} 

\vskip -.1in

\altaffilmark{5}{Physics Department, University of Notre Dame, IN
46556; pgarnavi@nd.edu}

\vskip -.1in

\altaffilmark{6}{Lawrence Livermore National Laboratory, L-43, 7000
East Avenue, Livermore, CA 94550, USA; mauche@cygnus.llnl.gov}

\vskip -.1in

\altaffilmark{7}{Mullard Radio Astronomy Observatory, Cavendish
Laboratory, Madingley Road, Cambridge CB3 0HE, England;
ggp1@cam.ac.uk}

\vskip -.1in

\altaffilmark{8}{Laboratory for High-Energy Astrophysics, NASA Goddard
Space Flight Center, Greenbelt, MD 20771; shrader@grossc.gsfc.nasa.gov}

\newpage

\normalsize

\section{INTRODUCTION}

X-ray novae (a.k.a. soft X-ray transients) are a type of X-ray binary
that typically remains quiescent for decades before brightening by as
much as $10^{7}$ in X-rays in a week. In outburst, the 2-10 keV X-ray
spectrum of most X-ray novae is dominated by thermal emission from the
inner accretion disk.  However, five X-ray novae are now known which
have failed to show this soft thermal component during outburst
(Brocksopp et al. 2001).  One of these five is XTE J1118+480. This
source is further distinguished by its strikingly low X-ray-to-optical
flux ratio (Hynes et al. 2000).

XTE J1118+480 was discovered on 2000 March 29 (Remillard et
al. 2000). During the month of March, {\it RXTE} All-Sky Monitor data show
that the X-ray intensity of the source increased steadily to
$\approx$~35 mCrab (2-12 keV).  Thereafter, the intensity remained
near that level for about three months before declining abruptly.  The
optical counterpart brightened from quiescence by about 6 mag to V
$\approx$~13 (Uemura et al. 2000).  Optical observations in outburst
and quiescence confirm that the orbital period is 4.08~hr (Patterson
et al. 2000; Uemura et al. 2000; McClintock et al. 2001; Wagner et
al. 2001).  A radio counterpart has also been observed (Pooley \&
Waldram 2000).  The most uncommon property of XTE~J1118+480 is its
exceptionally high galactic latitude, b~=~+62$^{\rm o}$, and its
correspondingly low reddening: E(B-V) $\approx$ 0.013 mag (N$_{H}
\approx 7.5 \times 10^{19}$~cm$^{-2}$; Hynes et
al. 2000). XTE~J1118+480 is the least reddened of all known X-ray
binaries.

Recently, dynamical measurements by two groups have established that
XTE~J1118+480 has a very large mass function, which sets a hard lower
limit of 6~M$_{\odot}$ on the mass of the compact X-ray source
(McClintock et al. 2001; Wagner et al. 2001).  Since this greatly
exceeds the maximum allowed stable mass of a neutron star in general
relativity (Rhoades \& Ruffini 1974), we refer to the compact primary
as a black hole.

The first epoch of an intensive multi-epoch, multiwavelength observing
campaign was reported by Hynes et al (2000).  Here we report on the
second epoch observations, centered on 2000 April 18, which are unique
in including a 0.24--7 keV grating spectrum obtained using the {\it Chandra
X-ray Observatory (CXO)}.  This spectrum is very important and
forces modification of the conclusions reached by Hynes et al. (2000).
Our near-simultaneous observations cover $\approx$ 80\% of the
electromagnetic spectrum from the infrared to hard X-rays; these
multiwavelength data are modeled in a companion paper by Esin et
al. (2001).

\section{NEAR-SIMULTANEOUS OBSERVATIONS}

A journal of the near-simultaneous observations that were made on or
near April 18 is given in Table 1. The April 18 {\it EUVE} data are
discussed extensively by Hynes et al. (2000), but they were not used
in their spectral energy distribution (SED).  The only data in common
between the SED presented by Hynes et al. and the SED presented herein
is the April 18 UKIRT data.  In the following we briefly discuss each
data set in turn.

{\it Rossi X-ray Timing Explorer (RXTE)} spectra were produced using
128-channel PCA data collected with PCUs 0, 2 and 3, and 64-channel
data collected with the HEXTE clusters A and B.  Following the
suggestions of Jahoda (2000), we have selected only the top layers of
each PCU and have combined the data for all three PCUs.  A response
matrix was generated using version 2.43 of {\it pcarsp}, and a
background estimate was generated using the 2000 January 31 blank sky
model for gain epoch 4 combined with version 2.1e of {\it pcabackest}.
A 1\% systematic error was added to the statistical error estimate and
each model was fit between 2.5--25 keV.  This systematic error was
required because the statistical uncertainty for the PCA is
significantly less than the uncertainty in the PCA response matrix,
which is estimated to be about 1\% (Jahoda 2000).  For the HEXTE we
used the standard response matrices and modeled the data between
15-200 keV, allowing for a constant normalization between the PCA data
and the data from the two HEXTE clusters. No systematic errors were
added to these data.  The background was estimated using the script
{\it hxtback} from FTOOLS version 5. The combined PCA and HEXTE data
are compatible with a power-law model for the emission.  The best-fit
photon index was $\Gamma = 1.782 \pm 0.005$ with a 2--200 keV flux of
$4.2\times10^{-9}$ ergs cm$^{-2}$ s$^{-1}$ and a $\chi^2$ of 164 for
139 dof. The photon index is consistent with the value measured 10
days earlier: $\Gamma = 1.8 \pm 0.1$ (Hynes et al. 2000).  A Gaussian
feature at 6.4 keV was evident in the residuals to the fit ($\chi^2
=$152 for 136 dof) with an equivalent width (EW) of 40~eV.  This
feature is very probably not due to the source for two reasons: First,
its strength is consistent with a feature seen in fits to the spectrum
of the Crab Nebula, where no Fe-K line is expected; thus it is
probably a systematic feature in the response matrices that we used
(Jahoda 2000).  Second, the feature does not appear in the {\it
Chandra} grating data described directly below.  For an assumed narrow
6.4~keV line, $\Delta$E/E = 0.04, the {\it Chandra} data imply a
3~$\sigma$ upper limit of EW~$<$~24~eV, which strongly rules against
the {\it RXTE} candidate Fe line.  For a broader assumed line width,
$\Delta$E/E = 0.10, the {\it Chandra} 3~$\sigma$ upper limit is
EW~$<$~38~eV, which marginally rules against the {\it RXTE} spectral
feature.  On the other hand, a very broad line, with a width
comparable to the energy resolution of the PCA detector
($\Delta$E/E~$\approx~0.2$) and an EW of 40~eV would have escaped
detection by {\it Chandra}.

{\it Chandra} observations were performed using the Low Energy
Transmission Grating (LETG) and the ACIS-S detector which yielded a
spectrum from 0.24 to 7 keV (1.8--52 \AA) with a resolution of about
0.04\AA.  The ``Level 1'' data were reduced using custom IDL
procedures in several steps: 1)~Sky coordinates were transformed to
grating coordinates by correcting for spacecraft roll; 2) the location
of zeroth order was determined by fitting a pair of 1-D Gaussian
profiles; 3) pulse height was converted to energy ($E_{PH}$) using
preflight gains and then corrected (node-by-node) to match the
energies inferred from the dispersion of the LETG; and 4) events were
selected using $r_l < E_{PH} / E_{LETG} < r_h$, where $r_l,r_h$ was
[0.70,1.15] for +1 events, [0.80, 1.20] for -1 events with $\lambda <
30$ \AA, and [0.65,1.25] for -1 events with $\lambda > 30$ \AA.  These
selections net $>$99\% of the observed counts.  All events within
2.5\arcsec\ of the dispersion line were included, except for the
events near detector gaps, which were ignored.  Any aperture size
between 1.5\arcsec\ and 3\arcsec\ would have yielded the same results.
No background was subtracted because the background rates were
negligible compared to the source rates ($<$ 1\%), except near the C-K
edge where the background rate was up to 5\% of the source rate.  The
events were binned and the effective areas were integrated over the
bins. After correcting for the fraction of counts dispersed by the
LETG fine support structure, we estimate that the fluxes are accurate
to $\sim$ 5\% in the 2-5 keV range.  XTE J1118+480 was the brightest
source observed with this grating and detector combination, so these
data are being used extensively to verify and update the instrumental
area, which could still have systematic errors of order 10-20\% in the
0.2-1.0 keV band (Marshall 2001). The largest uncertainties are in the
0.3-0.5 keV band.  The spectrum was searched at high resolution
(0.01\AA) for lines and edges with assumed widths comparable to the
instrumental resolution.  None was found apart from two weak,
unidentified lines, 13.24\AA\ (0.936 keV) and 9.36\AA\ (1.325 keV),
which were detected at significance levels of 5.6$\sigma$ and
4.6$\sigma$, respectively; the latter feature was detected only in the
third-order spectrum.  The probability of finding a feature in one bin
at $\geq 4.6\sigma$ among the 2500 bins searched is only 0.5\%. The
complete high-resolution spectrum is shown in Figure 1.  Given the
near absence of spectral features, the data were binned heavily on a
logarithmic grid ($\Delta$E/E = 0.02) to provide an accurate measure
of the continuum.  On time scales $\gtrsim$~1~ks, the total X-ray
intensity was constant to $\sim$~3\% during the 27 ks observation.
For the energy range 2--7 keV, the power-law photon index was
$\Gamma~=~1.77\pm0.04$, which is consistent with the {\it RXTE} values
quoted above.

{\it Extreme Ultraviolet Explorer (EUVE)} observations were performed
3 arcmin off-axis to extend the wavelength coverage of the {\it EUVE} short
wavelength spectrometer (SW) shortward of its normal cutoff (Marshall
et al. 1996).  Data from the SW spectrometer was binned in 2~\AA\
intervals and was corrected for IS absorption using the compilation of
H and He photoionization cross-sections of Rumph, Bowyer \& Vennes
(1994), with a mixture of neutral hydrogen to neutral and ionized
helium in the ratio 1:0.1:0.01.  Fortunately, the shape of the
extinction curve is largely independent of these details of the gas
mix. Flux calibration was achieved using the filter-corrected exposure
time and the off-axis effective area curve of H. Marshall et al. (in
preparation).  An independent analysis of these data is discussed by
Hynes et al. (2000); there are no significant differences between
their extracted spectrum and the one presented herein.

{\it Hubble Space Telescope (HST)} observations were made using the
Space Telescope Imaging Spectrograph (STIS) and the E140M, E230M,
G430L and G750L gratings.  An average calibrated spectrum for April 18
was constructed from standard {\it HST} pipeline data products.  The
spectrum exhibits emission lines of H$\alpha$ (weak), He\,{\sc ii}
(1640, 4686\,\AA), Si\,{\sc iv} (1394, 1403\,\AA) and N\,{\sc V}
(1239, 1243\,\AA), as well as Ly$\alpha$, higher order Balmer lines,
and the Balmer jump in absorption.  To identify the continuum spectrum
more clearly, and remove near-IR fringing, the spectra were rebinned
with the regions dominated by Ly$\alpha$ and N\,{\sc V} masked out.

The 3.8m United Kingdom InfraRed Telescope (UKIRT) was used to make
near--infrared photometric observations at JHKL'M' with IRCAM/TUFTI.
The measured magnitudes and comments on the data analysis can be found
in Hynes et al. (2000).

The Ryle Telescope was used to monitor the flux density at 15.2 GHz
using techniques similar to those described in Pooley \& Fender
(1997).  The phase calibrator used was J1110+440, and the flux-density
scale was established by observations of 3C48 and 3C286.

\section{SUPPORTING OBSERVATIONS}

The intensity of XTE J1118+480 was relatively constant during the
81~ks {\it EUVE} observation (Hynes et al. 2000) and during the 27 ks
{\it Chandra} observation (\S2).  In addition, the source was very stable on
time scales $\gtrsim$ hours at all wavelengths for weeks before and
after our April 18th observing campaign. To illustrate this constancy,
Figure~2 shows selected data for a $\sim 7$~week period around the
time of April~18, which is indicated by a dashed line in the
figure. Also, indicated by an arrow in Figure~2a is the time of the April 8
observing campaign, which occurred near the end of the rising phase of
the outburst (Hynes et al. 2000).

The 2--12 keV X-ray data shown in Figure 2a are consistent with a mean
intensity of I$_{\rm x}$ = 2.89 $\pm$ 0.41 counts $\rm s^{-1}$ (rms), which
corresponds to a relative intensity of 38 mCrab.  Optical photometric
data are shown in Figure 2b.  For these data we find mean magnitudes
of B = 13.02 $\pm$ 0.04 (rms) and I = 12.68 $\pm$ 0.05 (rms).  Figure
2c shows 15.2 GHz radio data with a mean flux density of S$_{\nu}$ =
8.68 $\pm$ 0.87 mJy (rms).  Thus the rms variability of the source
during this 7 week interval was 14\% in the X-ray, 5\% in the optical
and 10\% in the radio.  Moreover, much of this apparent variability is
due to measurement error.  The typical measurement error for the ASM
X-ray detectors was $\pm 10$\% (Fig. 2a) and for the Ryle Telescope it
was $\pm 5$\% (Fig. 2c).  Thus we conclude that for several weeks
around April 18, XTE J1118+480 was stable in intensity to better than
10\% on time scales of $\sim 1$ day in the radio, optical and X-ray
bands.

Because of the long-term stability of the source, we have summed 16
consecutive {\it RXTE} spectra obtained between April 13.39 and May 15.38 in
order to improve the counting statistics above 100 keV.  We used the
analysis techniques described in \S2.  These 16 observations can all
be fit individually by a power-law model with photon indices that
range between 1.77 and 1.81.  The summed spectrum has a total
integration time of 45.9~ks in the PCA and 16.8~ks in the HEXTE (only
cluster A was used).  A power-law model fit the summed spectrum with
$\Gamma = 1.779 \pm 0.003$ for a $\chi^{2}$ of 116 for 95 dof. The
best-fit power law that includes a simple exponential cutoff at high
energies, i.e. N(E)~$\propto$~E$^{-\Gamma}$e$^{-E/E_{c}}$, yielded a
cutoff energy of 940 keV for a $\chi^{2}$ of 103 for 94 dof.
Arbitrarily fixing the cutoff energy at 300 keV produced an
unacceptable fit with a $\chi^{2}$ of 150 for 95 dof. Of course, these
results do not rule against an abrupt, breaking cutoff at energies of
$\sim~150$~keV, since these energies are near the limit of HEXTE's
response.  Compared to the simple power-law model, a somewhat poorer
fit to the data was achieved using a comptonization model (``compTT'
in XSPEC; Arnaud \& Dorman 2000; Titarchuk 1994) with an electron temperature of 207
keV and an optical depth of 1.0 for a $\chi^{2}$ of 125 for 94 dof.

\section{THE SPECTRAL ENERGY DISTRIBUTION}

The most difficult problem in constructing the spectral energy
distribution (SED) is in properly correcting the EUV fluxes for IS
absorption.  It is not possible to obtain a precise measurement of
N$_{\rm H}$, the IS column depth.  For example, 21-cm measurements
imply N$_{\rm H}~= 1.34~\times~10^{20}$~cm$^{-2}$ (Dickey \& Lockman
1990).  On the other hand, the COBE maps of dust IR emission imply
N$_{\rm H} =~0.67~\times~10^{20}$~cm$^{-2}$ (Schlegel, Finkbeiner \&
Davis 1998).  However, neither of these values can be considered
secure, since N$_{\rm H}$ can vary by a factor $\sim$~2 on much
smaller angular scales than those probed by the surveys just mentioned
(Faison et al. 1998).  A line-of-sight estimate of N$_{\rm H}$ using
the Ca II lines implies a high value: N$_{\rm H}~=
2.8~\times~10^{20}$~cm$^{-2}$; however, the uncertainty in log(N$_{\rm
H}$) is at least 0.2 (Dubus et al. 2001). The source was too faint to
make a 21-cm absorption measurement feasible.

There is an additional complication with the IS absorption in the EUV:
A significant fraction of the absorption near 0.1~keV is attributable
to neutral and ionized helium; however, there are no existing data on
the absorbing columns for this line of sight.  Rigorous inclusion of
He absorption based on only the neutral hydrogen column would then
require knowledge of both the line of sight hydrogen and helium
ionization fractions, as well as the abundance of helium relative to
hydrogen.  Consequently, we estimate the IS absorption, parameterized
by N$_{\rm H}$, by assuming the neutral and ionized He number
densities relative to that of neutral hydrogen stated in $\S$2.  Given
all of these uncertainties, we, like Hynes et al. (2000), chose to
estimate N$_{\rm H}$ by examining the SED itself.

In Figure 3 we show four realizations of the SED for four assumed
values of N$_{\rm H}$.  Here we have omitted the infrared and radio
data. At energies above a few tenths of a keV, the spectrum can be
described in terms of a power law with varying spectral index as
follows.  At the highest energies, the photon index is $\approx 1.78$
(\S2 \& \S3).  At energies below $\sim 2$ keV, the spectrum becomes
harder and the photon index approaches $\approx$ 1.5.  It is possible
that the systematic errors discussed in \S2 contribute to this
hardening of the spectrum; however, it is unlikely that they can
account for all of it.

The only appreciable gap in the SED is centered at log($\nu$) $\approx
16.0$, where the ISM is opaque.  We now focus on the question of how
to connect up the {\it HST} data with the {\it EUVE} data across this unobserved
region.  It is apparent from Figure 3 that the dereddened {\it EUVE}
spectrum is very sensitive to the choice of N$_{\rm H}$ and is
consequently very uncertain.  Therefore, we first consider the
high-quality {\it HST} spectrum, which depends weakly on the choice of
reddening.  We can be confident that the {\it HST} spectrum is comprised of
substantial thermal emission because the Balmer jump in absorption is
apparent at log($\nu$) = 14.9. In the following we assume that the UV
emission is due chiefly to the accretion disk (see \S5).

We therefore consider the simple model spectrum of a steady accretion
disk as implemented in XSPEC (Arnaud \& Dorman 2000; Mitsuda et
al. 1984; Makishima et al. 1986). We do not include a power-law
component, which would increase the {\it EUVE} model fluxes somewhat.  These
multicolor disk blackbody spectra are meant to be illustrative; they
are not fits to the data.  Each disk spectrum shown in Figure 3 is
specified by two parameters: (1) the temperature at the inner edge of
the disk, T$_{in}$, and (2) the normalization, which we arbitrarily
take to be the dereddened {\it HST} flux at log($\nu$) = 15.1.  We first
consider the SED for N$_{\rm H}~= 1.6~\times~10^{20}$~cm$^{-2}$
(Fig. 3d). We find that no simple disk spectrum can match the steeply
rising {\it EUVE} spectrum, and the mismatch is only worse for higher values
of N$_{\rm H}$.  We therefore do not consider values of N$_{\rm
H}~\gtrsim~1.6~\times 10^{20}$~cm$^{-2}$ further.

We next examine N$_{\rm H}~= 0.75~\times~10^{20}$~cm$^{-2}$, which is
illustrated in Figure 3a.  By considering the {\it EUVE} data as part of the
nonthermal spectrum (see Hynes et al. 2000), it is possible to
accommodate a thermal disk component with kT$_{in}~\lesssim~12$~eV.
However, the comparison of the model to the data is somewhat
unsatisfying for two reasons. First, Hynes et al. made the reasonable
assumption that the EUV and hard X-ray fluxes were approximately
fitted by the same power law; however, the {\it Chandra} data complicate
this picture by revealing a wider range of power-law slopes.  Second,
the structure in the {\it EUVE} spectrum is not easily explained, although
it may possibly be attributable to the presence of a warm absorbing
medium around the source (see Esin et al. 2001).  In short, we
consider N$_{\rm H}~= 0.75 \times 10^{20}$~cm$^{-2}$ a viable choice;
however, we do not favor it.  Even lower values of N$_{\rm H}$ appear
less likely, since at long wavelengths the {\it EUVE} spectrum plunges
downward (Hynes et al. 2000) and has the wrong inflection to join up
with the model disk spectra.

We favor the two remaining spectra shown in Figures 3b \& 3c.  First
consider the case N$_{\rm H}~= 1.3~\times~10^{20}$~cm$^{-2}$
(Fig. 2c).  This case provides the closest match between one of the
model disk spectra (kT$_{in}$ = 24 eV) and the {\it EUVE} spectrum.  N$_{\rm
H}~= 1.3~\times~10^{20}$~cm$^{-2}$ probably corresponds to an upper
limit on the value of N$_{\rm H}$, since there may be a significant
non-disk component of UV emission at log($\nu$) = 15.1 (see \S5).  In
this case, the disk spectra in Figure 3c would need to be renormalized
downward, thereby creating some disagreement between the data and the
models: i.e. the {\it EUVE} data would then rise more steeply than the
models (cf. Fig. 3d).

Finally, consider the spectrum for N$_{\rm H}~=
1.0~\times~10^{20}$~cm$^{-2}$ (Fig. 3b).  At short wavelengths, the
{\it EUVE} spectrum conforms closely to the model for kT$_{in}$ = 22 eV,
whereas at longer wavelengths it straddles the models for kT$_{in}$ =
18-22 eV. If the normalization of the disk models were to be corrected
downward (as mentioned above), this model would provide a better
description of the data.  In short, we conclude that the most probable
value of the column density lies in the range N$_{\rm H}~=
1.0-1.3~\times~10^{20}$~cm$^{-2}$.  In the following, we adopt N$_{\rm
H}~= 1.3~\times~10^{20}$~cm$^{-2}$ and the version of the SED shown in
Figure 3c.  This spectrum, including the infrared and radio
data, is reproduced in more detail in Figure~4; see the caption for
further details about the individual data sets.  The data shown in
Figure~4 are available in digital form by request: For the {\it HST} data
contact C.~Haswell or R.~Hynes; for the remaining data contact
J.~McClintock.

\section{DISCUSSION}

Most black-hole X-ray novae in outburst reach the {\it high/soft} state (van
der Klis 1994) and spend considerable time there; consequently, most
X-ray novae are observed in this state.  The {\it high/soft} state is
dominated by a $\sim 1$~keV blackbody-like spectral component, which
is widely attributed to an accretion disk with its inner edge at or
near R$_{in}$ = 3R$_{\rm Schw}$ (R$_{\rm Schw}$~=~2GM/c$^{2}$), the
radius of the innermost stable orbit (Tanaka \& Lewin 1995).  A very
simple and successful model for this thermal disk spectrum is the
multicolor disk blackbody model (\S3; Mitsuda et al. 1984; Makishima
et al. 1986).

A few X-ray novae have been observed only in the {\it low/hard} state:
e.g. V404 Cyg, GRO J0422+32 and GS1354-64 in 1997 (Brocksopp et
al. 2001).  A number of canonical X-ray novae (e.g. Nova Mus 1991 and
GS2000+25) and Cyg X-1 have also been observed in the {\it low/hard} state
(Tanaka \& Lewin 1995).  For the sources observed in this state, the
non-stellar optical/UV spectrum has provided evidence for the presence
of an outer accretion disk. However, for these sources it has not been
possible to determine directly the properties of their inner disks,
which emit in the EUV and soft X-ray bands, due to the high IS column
depths: N$_{\rm H}~\gtrsim~10^{21}$~cm$^{-2}$.  Observations of XTE
J1118+480 with N$_{\rm~H}~\approx~1.3~\times~10^{20}$~cm$^{-2}$
provide the first detection of radiation from the inner accretion disk
for an X-ray nova in the {\it low/hard} state.

A simple explanation for the low thermal temperature in XTE J1118+480
is that the inner edge of the accretion disk is quite far out compared
to R$_{in}~=~3$R$_{\rm Schw}$.  This is the expected state of affairs
in some models of the {\it low/hard} state (e.g. Esin, McClintock \& Narayan
1997).  Moreover, there is good indirect evidence based on Compton
reflection models that such large, cool disks do exist in the {\it low/hard}
state of several systems (e.g., see Gierlinski et al. 1997; Miller et
al. 2001).

Based on our adopted value of N$_{\rm H}$ (\S4) and a distance of 1.8
kpc (McClintock et al. 2001; Wagner et al. 2001), we can make a rough
estimate of the inner disk radius for kT$_{in} = 24$ eV (Fig. 3c) as
follows.  For an intensity of 38 mCrab and a photon spectral index of
1.78 (\S3), the 1--160 keV X-ray luminosity is L$_{x} \approx 1.2
\times 10^{36}$~ergs~s$^{-1}$.  For an assumed value of the accretion
efficiency, $\epsilon$, the X-ray luminosity determines the mass
accretion rate, \.{M} = L$_{x}$/$\epsilon$c$^{2}$, and hence the inner
radius of a steady-state disk:
R$_{in}$~=~(3GM$_{x}$\.{M}/8$\pi\sigma$T$_{in}^{4}$)$^{1/3}$
(e.g. Frank, King \& Raine 1992).  For a canonical efficiency of 10\%,
we find R$_{in} = 60$R$_{\rm Schw}$, and for advection-dominated flow
with an assumed efficiency of 0.1\% we find R$_{in} = 285$R$_{\rm
Schw}$.  A comparable value for R$_{in}$ can be obtained by using the
normalization constant determined for the multicolor disk blackbody
model, which depends only on the solid angle subtended by the inner
disk and the inclination of the disk (Arnaud \& Dorman 2000).
Assuming that the flux at log($\nu$) = 15.1 is due solely to the disk
(see below) and assuming a high inclination, i = $80^{\rm~o}$ (Wagner
et al. 2001), we find R$_{in} = 34$R$_{\rm Schw}$ for the model with
kT = 24 eV shown in Figure 3c.

We note that the radio/IR data and the {\it HST} data below the Balmer jump
can all be reasonably well fit by a single power law, $\nu$F$_{\nu}
\propto~\nu^{1.10}$, which steepens at longer radio wavelengths (Hynes
et al. 2000). Thus, in the IR/radio band there is an additional
component of emission above that expected for a multicolor disk blackbody
spectrum (Hynes et al. 2000; Esin et al. 2001).  This component may be
associated with an outflow from the system (Esin et al. 2001; Fender
2001).

Could this IR/radio component affect our earlier conclusions on the
temperature of the thermal accretion disk component?  In \S4 we
assumed that the UV flux is dominated by the spectrum of the accretion
disk.  This assumption is supported by the presence of a
$\approx~0.09$ mag Balmer jump in the spectrum.  Unfortunately, the
strength of the Balmer jump in absorption cannot be used to determine
quantitatively the fraction of the total flux that is due to the
accretion disk.  In a disk there are several conditions that can
reduce or even eliminate the Balmer jump: e.g. high inclination and
X-ray heating (la Dous 1989).  Indeed, Balmer jumps are often absent
or appear in emission in the spectra of dwarf novae (la Dous 1989;
Williams 1983) and one was not detected in the spectrum of X-ray Nova
Mus 1991 during outburst (Cheng et al. 1992).  In short, we cannot
rule out the hypothesis that the IR/radio component contributes to the
UV spectrum of XTE~1118+480.  However, the presence of a significant
Balmer jump argues for appreciable disk emission.  In this work, we
assume that the disk flux is dominant in the UV band.

In conclusion, the global spectrum consists of three components: (1) a
$\approx$~24~eV thermal component due to the accretion disk with a
large inner radius ($\gtrsim 35$R$_{\rm Schw}$); (2) a quasi power-law
component, which extends from $\sim 0.4$ keV to at least 160 keV; and
(3) a third component that dominates the spectrum at wavelengths
greater than several microns.  A model for the first two components
and a discussion of the third one are presented in a companion paper
by Esin et al. (2001).

\bigskip

We thank the {\it CXO} Director H. Tananbaum for granting Director's
Discretionary Time, the entire {\it Chandra} team for their superb effort and
enthusiasm, and NASA for providing support for J.E.M., M.R.G., and
J.J.D. through grant DD0-1003X and contract NAS8-39073.  The {\it
EUVE\/} observations were made possible by a generous grant of
Director's Discretionary Time by {\it EUVE\/} Project Manager
R.~Malina, the efforts of {\it EUVE\/} Science Planner M.~Eckert, the
staff of the {\it EUVE\/} Science Operations Center at CEA, and the
Flight Operations Team at Goddard Space Flight Center.  This work
includes observations with the NASA/ESA {\it Hubble Space Telescope},
obtained at the Space Telescope Science Institute, operated by the
Association of Universities for Research in Astronomy, Inc.\ under
NASA contract No.\ NAS5-26555.  We would like to thank the {\it HST}
and {\it RXTE} support staff for ongoing efficient support.  UKIRT is
operated by the Joint Astronomy Centre on behalf of the U.K. Particle
Physics and Astronomy Research Council.  C.A.H., R.I.H. and
S.C. acknowledge support from grant F/00-180/A from the Leverhulme
Trust.  J.E.M. acknowledges helpful discussions with A. Siemiginowska
and N. Brickhouse.  We thank an anonymous referee for several
constructive comments.

\newpage

\footnotesize
\begin{center}
\begin{tabular}{llccll}
\multicolumn{6}{c}{TABLE 1} \\
\multicolumn{6}{c}{OBSERVATIONS ON OR NEAR 2000 APRIL 18 UT} \\
\hline\hline
&&&&&Net Obs. \\
Observatory&Instrument&Bandpass&log($\nu$)&Observation Interval (UT)&Time (ks) 
\\ \hline
{\it RXTE}&HEXTE&15--200 keV&18.56-19.68&18 Apr 19:28--18 Apr 23:00&1.1
 \\
&PCA&2.5--25 keV&17.78-18.78&18 Apr 19:28--18 Apr 23:00&2.8 \\
{\it Chandra}&LETG/ACIS-S&0.24--7 keV&16.76-18.23&18 Apr 18:16--19 Apr 02:16&27.2 \\
{\it EUVE}&SW&0.10--0.17 keV&16.38-16.61&16 Apr 21:34--19 Apr 14:28&81.1 \\
{\it HST}&STIS&1155--10250\AA&14.47-15.41&18 Apr 13:40--18 Apr 17:44&6.2 \\
UKIRT&IRCAM/TUFTI &1-5 $\mu$&13.78-14.48&18 Apr 12:00&0.01--0.06 \\
Ryle Telescope&15.2 GHz receiver&$2.0$ cm&10.18&18Apr
17:11--18 Apr 18:36& 5.1 \\ \hline
\end{tabular}
\end{center}

\newpage \figcaption[fig1.eps] {First-order ACIS/LETG count spectrum
binned at 0.025\AA\, as obtained in a 27~ks observation with {\it Chandra}.
The dashed line corresponds to a power-law model spectrum folded
through the instrument response with $\Gamma~ =~1.78$ (\S3) and
N$_{\rm H}~= 1.3~\times~10^{20}$~cm$^{-2}$.  Apart from a weak,
unidentified feature centered at 13.24\AA\ (\S2), this first-order
spectrum is devoid of lines.}

\figcaption[fig2.eps] {A seven week record of the intensity of XTE
J1118+480 measured at X-ray, optical and radio frequencies.  The
dashed line corresponds to 2000 April 18.5 UT, the nominal time of the
intensive observing campaign reported on herein. The arrow is drawn at
2000 April 8.5 UT, the nominal time of the observations conducted by
Hynes et al. (2000).  (a) A 2--12 keV {\it RXTE} ASM light curve.  The data
shown are daily-average intensities, which were obtained from the MIT
ASM web page. (b) UBVRI optical photometric data were obtained on 17
occasions using the 1.2m telescope at the F. L. Whipple Observatory.
Only the B and I light curves are shown here.  (c) A 15.2 GHz radio
light curve obtained using the Ryle Telescope.}

\figcaption[fig3.eps] {Four realizations of the SED for the indicated
values of the column density.  An inspection of the panels (a) -- (d)
shows that the {\it HST} UV spectrum and the low-energy portion of the {\it Chandra}
X-ray spectrum are only mildly sensitive to the choice of N$_{\rm H}$.
However, the {\it EUVE} spectrum is extremely sensitive.  The solid curves
represent multicolor disk blackbody model spectra, which have been
normalized to the {\it HST} flux at log($\nu$) = 15.1. The models are
specified by the temperature at the inner edge of the accretion disk,
T$_{in}$.  The four models in each panel are equally spaced by 2 eV
and the lowest and highest values of kT$_{in}$ are indicated in the
figure.}

\figcaption[fig4.eps] {A blowup of the SED corrected for the adopted
column depth of N$_{\rm H}~= 1.3~\times~10^{20}$~cm$^{-2}$. This is
the same SED shown in Figure 3c, except that the infrared data are
included here. The inset shows all of the same data plus the 15.2 GHz
radio data for April 18.  The following comments about the individual
data sets apply also to Figure~3: (1) The UKIRT flux densities
reported by Hynes et al. (2000) have been multiplied by 0.8 to force
them to approximately match the {\it HST} spectrum; the discrepancy may be
due to the short UKIRT exposures and the rapid optical variability of
the source (Patterson 2000).  (2) The {\it HST} data were dereddened
(Cardelli, Clayton \& Mathis 1989; Predehl \& Schmitt 1995). No ad hoc
corrections have been applied to these data; error bars are plotted on
every eighth point.  (3) The {\it EUVE} data become quite uncertain near the
lowest energy (0.10 keV); error bars are displayed on approximately
every third data point. (4) The {\it Chandra} LETG spectrum has very small
statistical errors; error bars are drawn on every tenth data
point. The spectrum shown here is truncated at 6 keV; otherwise, no ad
hoc corrections have been made to this spectrum.  (5) The {\it RXTE} data
begin at 3 keV and thus overlap substantially with the LETG spectrum.
The small filled circles without error bars show the spectrum for
April 18.  The summed spectrum (\S3) is plotted as open circles with
error bars.  The statistical significance of the last two data points
in the summed spectrum are 3.8$\sigma$ at 154 keV and 2.8$\sigma$ at
162 keV. Since Crab observations indicate that {\it RXTE} systematically
produces fluxes that are high by about 15\% and since some modest
variability is expected, the fluxes for April 18 were multiplied by
0.84 to match the overlapping LETG fluxes; similarly, the
summed-spectrum fluxes were multiplied by 0.91.}

\newpage
\begin{figure}
\figurenum{1}
\plotone{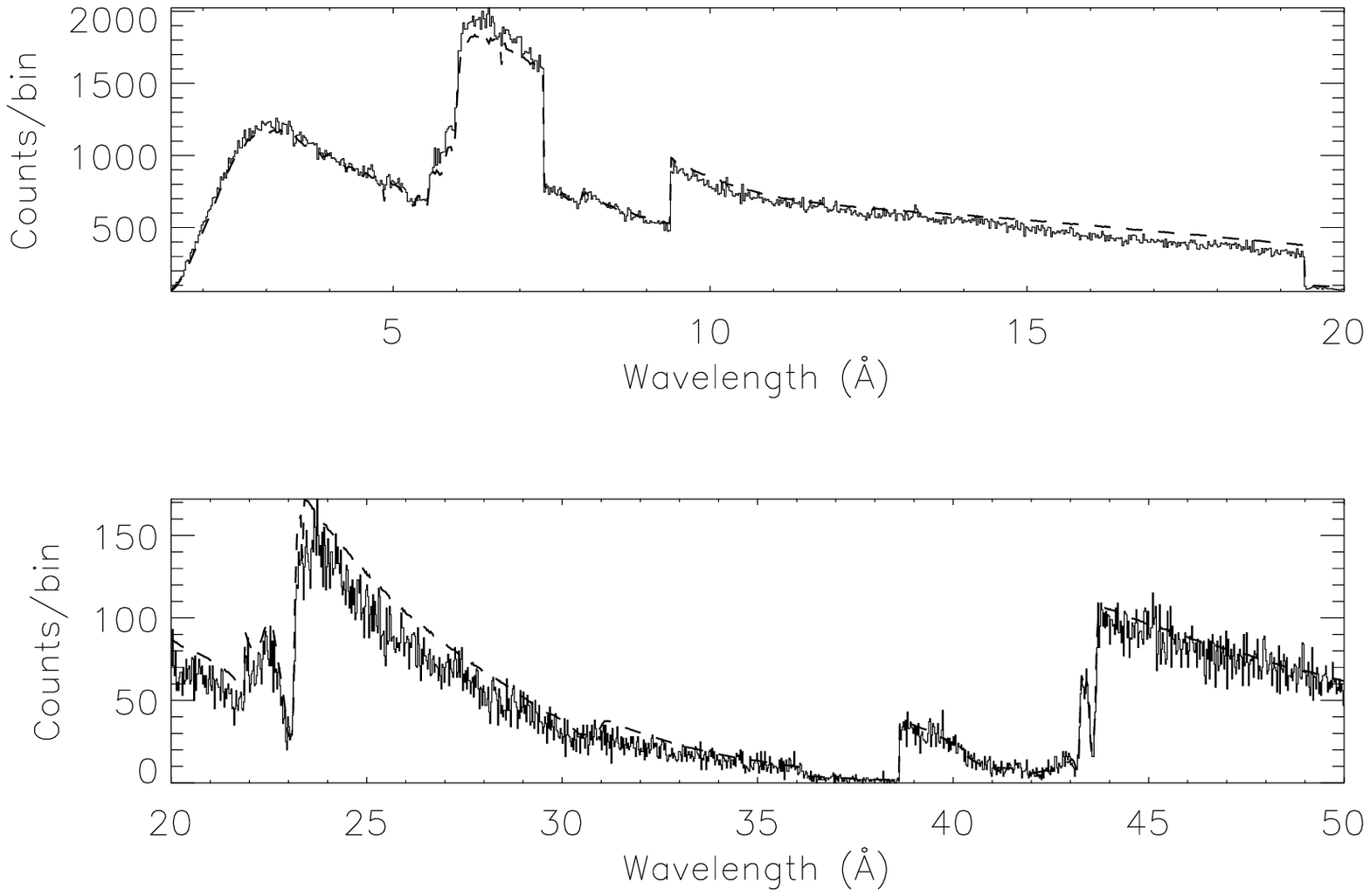}
\caption{ }
\end{figure}

\newpage
\begin{figure}
\figurenum{2}
\plotone{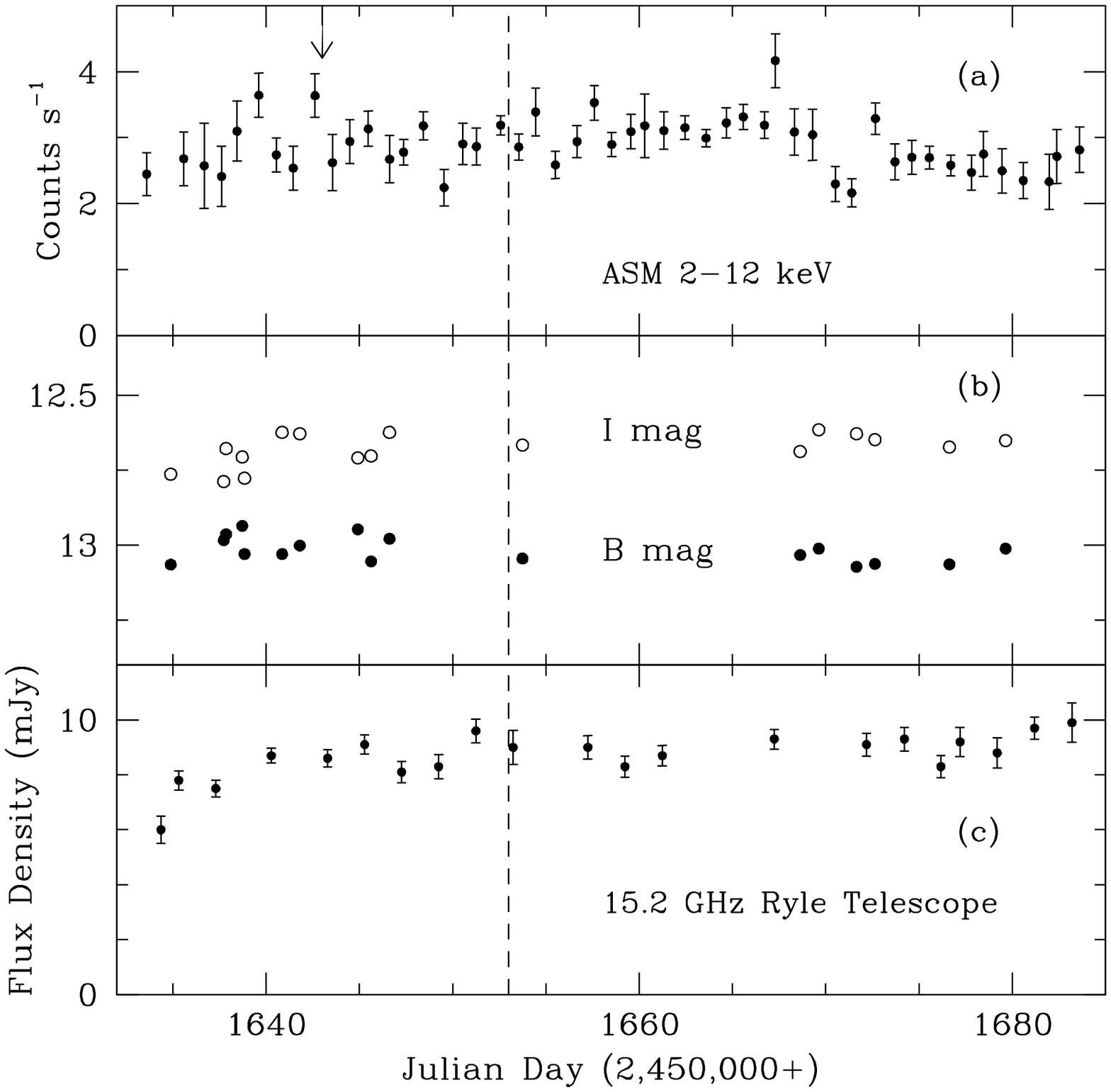}
\caption{ }
\end{figure}

\newpage
\begin{figure}
\figurenum{3}
\plotone{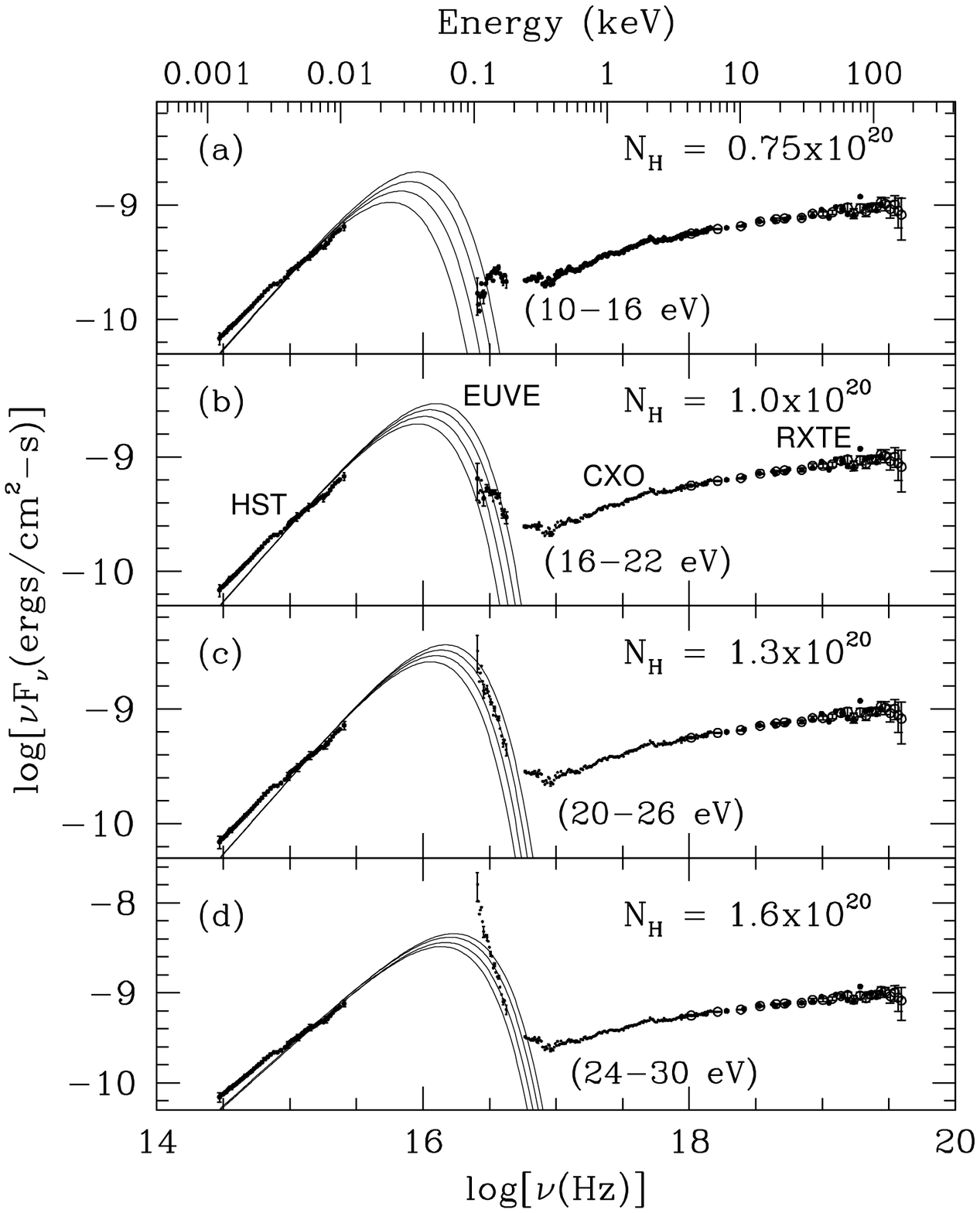}
\caption{ }
\end{figure}

\newpage
\begin{figure}
\figurenum{4}
\plotone{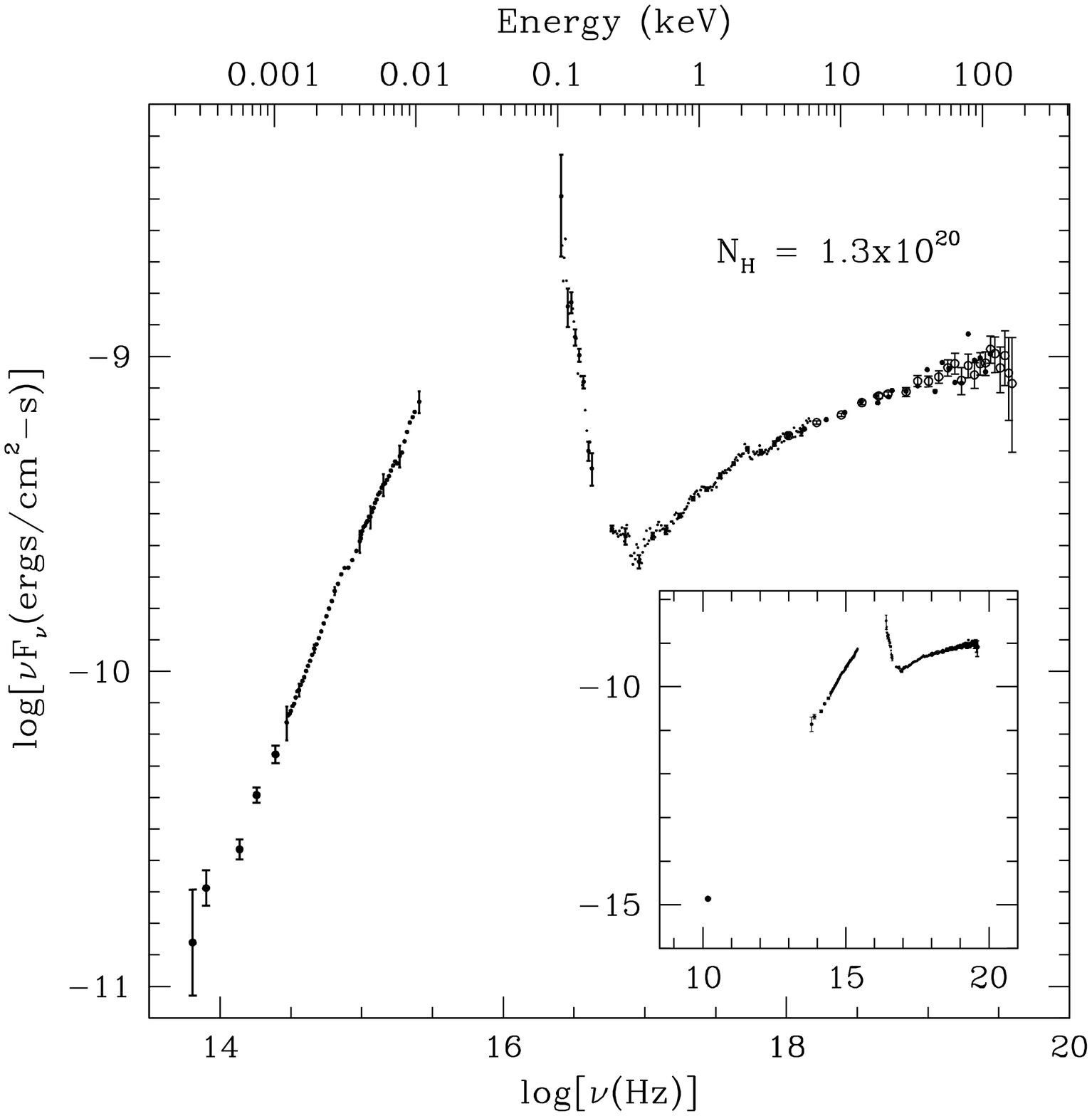}
\caption{ }
\end{figure}

\end{document}